\documentclass[aps,prseries,twocolumn,amsmath,amssymb,superscriptaddress,floatfix]{revtex4-2}
\usepackage{amsmath}
\usepackage{subfigure}
\usepackage{graphicx}
\usepackage{xcolor}
\usepackage{bm}
\usepackage[colorlinks=true,linkcolor=blue,citecolor=blue,urlcolor=blue,]{hyperref}
\usepackage[doipre={doi:~}]{uri}
\usepackage{lineno}
\graphicspath{{./figures/}}
\usepackage{ulem}

\begin{document}

\title{Density-wave-like gap evolution in La$_3$Ni$_2$O$_7$ under high pressure \\revealed by ultrafast optical spectroscopy}

\author{Yanghao Meng$^{1,2,*}$, Yi Yang$^{1,3,*}$, Hualei Sun$^{4,*}$, Sasa Zhang$^{3,5}$, Jianlin Luo$^{1,2}$, Liucheng Chen$^{1,2,6}$, Xiaoli Ma$^{1,2,6}$, Meng Wang$^{7,\dag}$, Fang Hong$^{1,2,6,\dag}$, Xinbo Wang$^{1,2,\dag}$, and Xiaohui Yu$^{1,2,6,\dag}$}

\affiliation{
\\$^1$Beijing National Laboratory for Condensed Matter Physics, Institute of Physics, Chinese Academy of Sciences, Beijing 100190, China
\\$^2$School of Physical Sciences, University of Chinese Academy of Sciences, Beijing 100190, China
\\$^3$Key Laboratory of Education Ministry for Laser and Infrared System Integration Technology, Shandong University, Qingdao 266237, China
\\$^4$School of Science, Sun Yat-sen University, Shenzhen 518107, China
\\$^5$School of Information Science and Engineering, Shandong University, Qingdao 266237, China
\\$^6$Songshan Lake Materials Laboratory, Dongguan, Guangdong 523808, China
\\$^7$Center for Neutron Science and Technology, Guangdong Provincial Key Laboratory of Magnetoelectric Physics and Devices, School of Physics, Sun Yat-Sen University, Guangzhou, China
\\$^*$These people contribute equally to the present work.
\\$^{\dag}$Corresponding authors: wangmeng5@mail.sysu.edu.cn, hongfang@iphy.ac.cn, xinbowang@iphy.ac.cn, yuxh@iphy.ac.cn.
}



\begin{abstract}
Density wave (DW) order is believed to be correlated with superconductivity in the recently discovered high-temperature superconductor La$_3$Ni$_2$O$_7$. However, experimental investigations of its evolution under high pressure are still lacking. Here, we explore the quasiparticle dynamics in bilayer nickelate La$_3$Ni$_2$O$_7$ single crystals using ultrafast optical pump-probe spectroscopy under high pressures up to 34.2 GPa. At ambient pressure, the temperature-dependent relaxation dynamics demonstrate a phonon bottleneck effect due to the opening of an energy gap around 151 K. The	energy scale of the DW-like gap is determined to be 66 meV by the Rothwarf-Taylor model. Combined with recent experiential results, we propose that this DW-like transition at ambient pressure and low temperature is spin density wave (SDW). With increasing pressure, this SDW order is significantly suppressed up to 13.3 GPa before it completely disappears around 26 GPa. Remarkably, at pressures above 29.4 GPa, we observe the emergence of another DW-like order with a transition temperature of approximately 135 K, which is probably related to the predicted charge density wave (CDW) order. Our study provides the experimental evidence of the evolution of the DW-like gap under high pressure, offering critical insights into the correlation between DW order and superconductivity in La$_3$Ni$_2$O$_7$. 
\end{abstract}
\maketitle
~\\
\noindent{\bf\large Introduction}\\

Nickel-based superconductors have attracted significant attention since the first member $\mathrm{Nd_{0.8}Sr_{0.2}NiO_2}$ was discovered\cite{li2019superconductivity,Zeng2022Supercond,osada2020phase,osada2021nickelate}. They have similar $d$ electron configurations resembling cuprates, suggesting the potential high-temperature superconductivity. This hypothesis was further supported by recent findings, where La$_3$Ni$_2$O$_7$ single crystal was found to show a superconducting transition with  $T_c \approx 80$ K at pressures above 14 GPa\cite{sun2023signatures}.  Many experiments reported its superconducting properties at high pressures\cite{zhang2023high,zhou2023evidence,puphal2023unconventional,PhysRevX.14.011040}. However, the mechanism of its superconductivity is still unclear and under debate.\cite{jiang2023high, PhysRevLett.131.126001, PhysRevLett.131.206501, PhysRevB.108.L201121,gu2023effective,chen2023critical,PhysRevB.108.125105,lu2023interplay,lange2023feshbach, PhysRevB.108.L180510,yi2024antiferromagnetic,PhysRevB.108.L140504,shen2023effective}.

The interplay between density wave order and superconductivity is a widely investigated topic in high-temperature superconductors since they are expected  to be strongly related\cite{RevModPhys.87.457}. In La$_3$Ni$_2$O$_7$, two DW transitions have been proposed at ambient pressure when the temperature is decreased\cite{puphal2023unconventional,chen2023evidence,dan2024spin,wu2001magnetic,liu2022evidence,kakoi2023multiband,zhou2024electronic,khasanov2024pressure,PhysRevX.14.011040}.  Various measurements, including resonant inelastic X-ray scattering (RIXS)\cite{zhou2024electronic}, muon spin rotation ($\mu$SR) experiments\cite{khasanov2024pressure,chen2023evidence}, Nuclear magnetic resonance (NMR)\cite{kakoi2023multiband, dan2024spin} have identified the presence of SDW transition around 150 K. While the CDW transition was suggested in either the transport or optical conductivity measurements, with transition temperature varying from 110 to 130 K\cite{sun2023signatures,kakoi2023multiband,liu2022evidence,PhysRevX.14.011040,liu2023electronic}. The complex DW behaviors in $\mathrm{La_3Ni_2O_7}$, which was proposed to stem from the scattering between the multiple Fermi surface sheet contributed mainly by the two $e_g$ Ni $d_{x^2-y^2}$ and $d_{z^2}$ orbitals\cite{PhysRevLett.131.126001, PhysRevLett.131.206501, PhysRevB.108.L201121,PhysRevLett.132.146002,PhysRevB.108.125105,PhysRevB.108.L140505,PhysRevLett.131.236002,PhysRevLett.132.036502}, are significantly influenced by the temperature and pressure\cite{zhang2023high,liu2022evidence,khasanov2024pressure}.  Thus, a thoroughly  investigation of the evolution of the DW orders under pressure is crucial for unraveling the pairing mechanism of superconductivity in this nickelate.

\begin{figure*}[tbp]
	\centering
	\includegraphics[width=6 in]{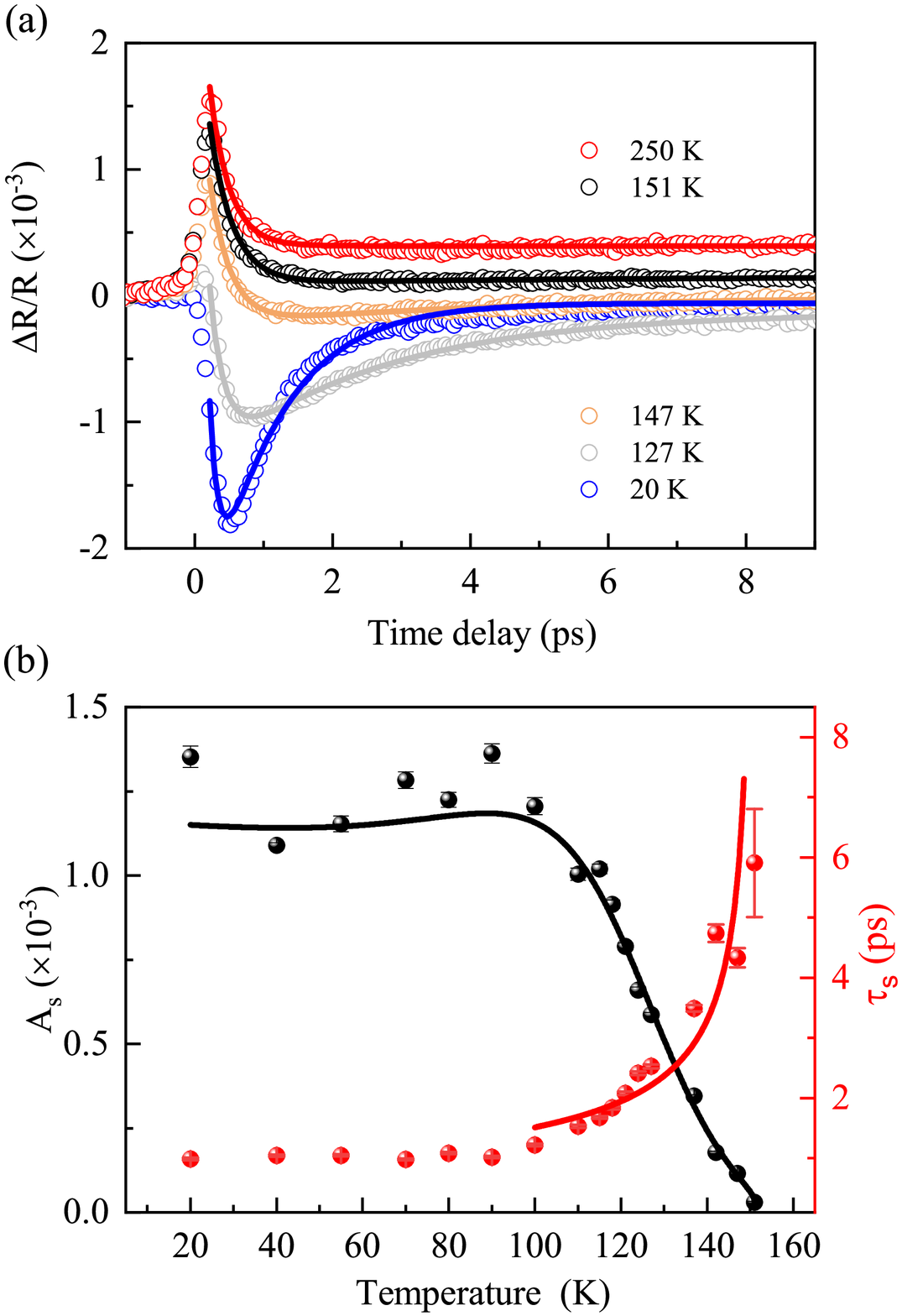}
	\subfigure{\label{1a}}
	\subfigure{\label{1b}}
	\caption{\textbf{Pump-probe spectra and the extracted parameters near ambient pressure.} (a) $\Delta R / R$ signals at several selected temperatures near ambient pressure. The experiential data can be well fitted by one and two exponential decays above and below 151 K, respectively. The solid lines are the fitting curves. (b) Temperature dependent amplitude $\mathrm{A_s}$ and relaxation time $\tau_{\mathrm{s}}$. $\mathrm{A_s}$ decreases to nearly zero around 151 K, where $\tau_{\mathrm{s}}$ shows a clear divergence. The solid lines are fitting results according to RT model. Error bars are the standard error in the exponential fitting.}
	\label{1}
\end{figure*}

Studies of the DW orders in La$_3$Ni$_2$O$_7$ at high pressure are currently insufficient due to the limited availability of experimental tools for reliable high-pressure measurements. Currently, most of the reported experimental results of  La$_3$Ni$_2$O$_7$ are focused on the transport properties. However, the signature of DW transition in the transport measurements usually becomes indistinguishable starting from 3 GPa\cite{liu2022evidence,wu2001magnetic,zhang2023high,PhysRevX.14.011040}. Recently, there have been a few tentative works to study the DW order at low pressure. For example, a $\mu$SR experiment\cite{khasanov2024pressure} indicates the DW order can persist at least to 2.3 GPa with an enhanced DW transition temperature, but the data under pressure higher than 2.3 GPa is still lacking. On the other hand, the superconducting volume fraction is relatively low and exhibits sample dependence, indicating that the currently available samples are highly inhomogeneous, as evidenced by X-ray diffraction \cite{puphal2023unconventional,chen2024polymorphism,wang2024long} and scanning transmission electron microscopy measurements\cite{puphal2023unconventional,dong2023visualization}. The crystal imperfections, such as oxygen vacancies, and the existence of multiple structural phases may obscure the intrinsic properties of the correct phase responsible for superconductivity. Up to now, how the DW orders evolve under high pressure remains unknown and requires further investigation.

Here, we report the evolution of DW-like orders in La$_3$Ni$_2$O$_7$ under high pressure using ultrafast optical spectroscopy. Time-resolved optical pump-probe spectroscopy has been widely employed to study nonequilibrium quasiparticle dynamics in various materials exhibiting superconductivity and density wave phenomena, due to its extreme sensitivity to the presence of energy gap\cite{giannetti2016ultrafast,dong2023recent}. However, performing pump-probe experiments under high pressure and low temperature is challenging due to the technical difficulties in combining high-pressure equipment with cryogenic systems while maintaining optical access for ultrafast laser pulses\cite{Wu2021On}. Despite these challenges, such experiments provide valuable insights into the behavior of materials under extreme conditions\cite{PhysRevLett.126.027402,PhysRevB.108.035101,PhysRevB.109.064307}. In this work, we observed the DW gap opening near ambient pressure below a transition temperature $T_{\mathrm{DW}}$, as evidenced by the phonon bottleneck (PB) effect. The gap fitted by Rothwarf-Taylor (RT) model is $\Delta_{\mathrm{DW}}$ = 66 meV. As the pressure is increased from 0 to 13.3 GPa, the low-pressure DW order is significantly suppressed with decreasing $T_{\mathrm{DW}}$. The PB effect is relatively weak at higher pressures and the energy gap decreases slightly with increasing pressure before becoming indistinguishable around 26 GPa. Above 29.4 GPa, another DW phase appears, indicated by the re-emergence of the PB effect and a drastic increase in the transition temperature. Our results report a thorough evolution of the complex DW orders under high pressure, providing key experimental information for understanding the mechanism of superconductivity in nickelate.

		\begin{figure*}[!htbp]
		\centering
		\includegraphics[width=\textwidth]{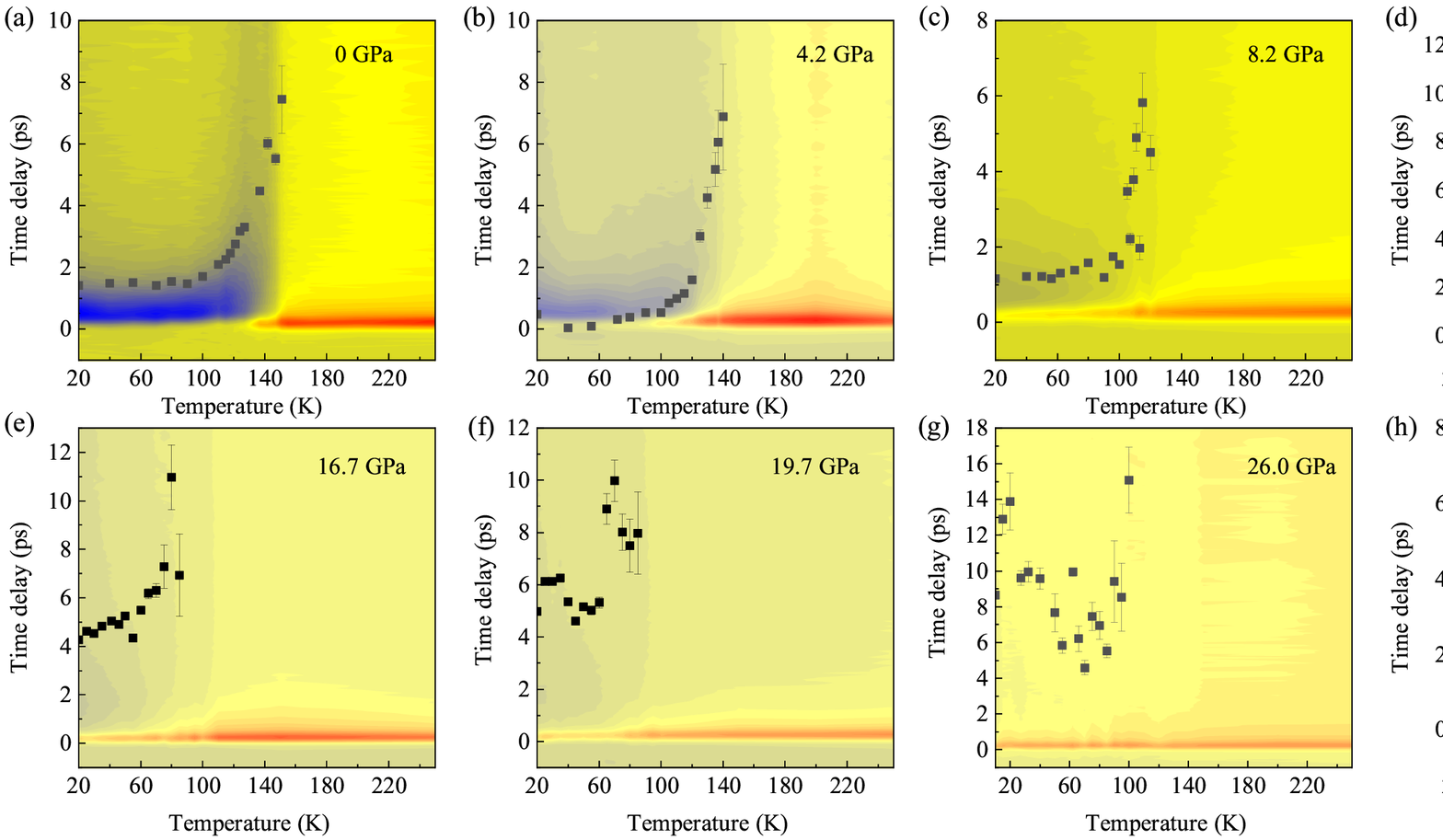}
		\subfigure{\label{2a}}
		\subfigure{\label{2b}}
		\subfigure{\label{2c}}
		\subfigure{\label{2d}}
		\subfigure{\label{2e}}
		\subfigure{\label{2f}}
		\subfigure{\label{2g}}
		\subfigure{\label{2h}}
		\caption{\textbf{Temperature dependent pump-probe spectra measured at different pressures.} (a) 0 GPa, (b) 4.2 GPa, (c) 8.2 GPa, (d) 13.3 GPa, (e) 16.7 GPa, (f) 19.7 GPa, (g) 26 GPa and (h) 34.2 GPa. The negative component exists at low temperature, and vanishes at temperature higher than $T_{\mathrm{DW}}$ for all pressures. The scatters in each panel are the extracted $\tau_{\mathrm{s}}$. Phonon bottleneck effect are clearly observed except for 26 GPa, indicating the suppression of DW orders. Error bars indicate the standard error in the exponential fitting.}
		\label{2}
	\end{figure*}

~\\
\noindent{\bf\large Results}\\

\noindent{\bf DW gap opening near ambient pressure}

Figure \ref{1a} shows the time-resolved reflectivity change $\Delta R / R$ in La$_3$Ni$_2$O$_7$ at several selected temperatures near ambient pressure. At high temperature, photoexcitation leads to a quick rise in the reflectivity, followed by a fast decay into a constant offset. The relaxation time exhibits minor variations as the temperature increases to 250 K. Below 151 K, an additional long-lived component with a negative amplitude appears, which relaxes more quickly and increases in amplitude as the temperature decreases further. This transition, where the initial positive change in $\Delta R / R$ turns negative, corresponds to the expected DW-like transition at ambient pressure. Consequently, we attribute the fast decay signal to the electron-phonon thermalization and the slow-decay component to the recombination across the DW gap, as discussed in detail below. Accordingly, we fit the data using a single-component exponential function, $\Delta R / R= \mathrm{A_f} e^{-t / \tau_{\mathrm{f}}}+C$ above $T_{\mathrm{DW}}$, and two-component decay function, $\Delta R / R=\mathrm{A_f} e^{-t / \tau_{\mathrm{f}}}-{\mathrm{A_s}} e^{-t / \tau_{\mathrm{s}}}+C$ at low temperature, where $\mathrm{A}$ and $\tau$ represent the relaxation amplitude and decay time, respectively (Supplementary Note 1). The subscripts ($\mathrm{f}$ and $\mathrm{s}$) denote the fast and slow relaxation processes, respectively. $C$ is a constant offset. The experimental data can be fitted quite well as shown in Fig. \ref{1a}. The extracted $\mathrm{A_s}$ and $\tau_{\mathrm{s}}$ as a function of temperature are depicted in Fig. \ref{1b}. Below $T_{\mathrm{DW}}, A_s$ increases sharply from zero, while $\tau_{\mathrm{s}}$ shows a continuous divergence. Our subsequent analysis suggests that the anomalous behavior around $T_{\mathrm{DW}}$ can be explained by a relaxation bottleneck associated with the opening of a DW-like gap.

To explain the slow relaxation process in $\mathrm{La_3Ni_2O_7}$, we employ the RT model\cite{rothwarf1967theory}. It is a phenomenological model that was initially proposed to describe the relaxation of photoexcited carriers in superconductors where the formation of a gap in the electronic density of states leads to a relaxation bottleneck. When the energy gap is comparable to the phonon energy, the phonons emitted during quasiparticle relaxation can re-excite the quasiparticles, thereby impeding their relaxation back to equilibrium. The RT model has also been shown to be applicable to other systems with gap opening in the density of states, such as charge/spin density wave, and heavy fermion materials\cite{giannetti2016ultrafast,dong2023recent}. Based on this model, the thermally quasiparticle density $n_T$ is related to the transient reflectivity amplitude $A$ via $n_T \propto[A(T) / A(T \rightarrow 0)]^{-1}-1$. Combining the relationship of $n_T \propto \sqrt{\Delta(T) T} \exp [-\Delta(T) / T]$, we obtain\cite{PhysRevB.59.1497}:
\vspace{-1em}

\begin{equation}
	A(T) \propto \frac{\Phi /\left(\Delta(T)+k_B T / 2\right)}{1+\gamma \sqrt{2 k_B T / \Delta(T)} \exp \left[-\Delta(T) / k_B T\right]}
\end{equation}

\noindent where the $\Phi$ is the pump fluence, $\Delta(T)$ is the temperature dependent gap energy, $k_B$ is the Boltzmann constant, and $\gamma$ is a fitting parameter. In the RT model, the relaxation time near transition temperature is dominated by phonons with frequency \textbf{ $\hbar\omega \geq 2 \Delta$ }transferring their energy to lower frequency phonons with $\hbar\omega<2 \Delta$, so the re-excitation of the condensed quasiparticles would stop. The relaxation time $\tau$ near transition temperature is given by\cite{PhysRevB.59.1497} :
\vspace{-2em}

\begin{equation}
\tau^{-1}(T) \propto \Delta(T),
\end{equation}

\begin{figure*}[tbp]
	\centering
	\includegraphics[width=6in]{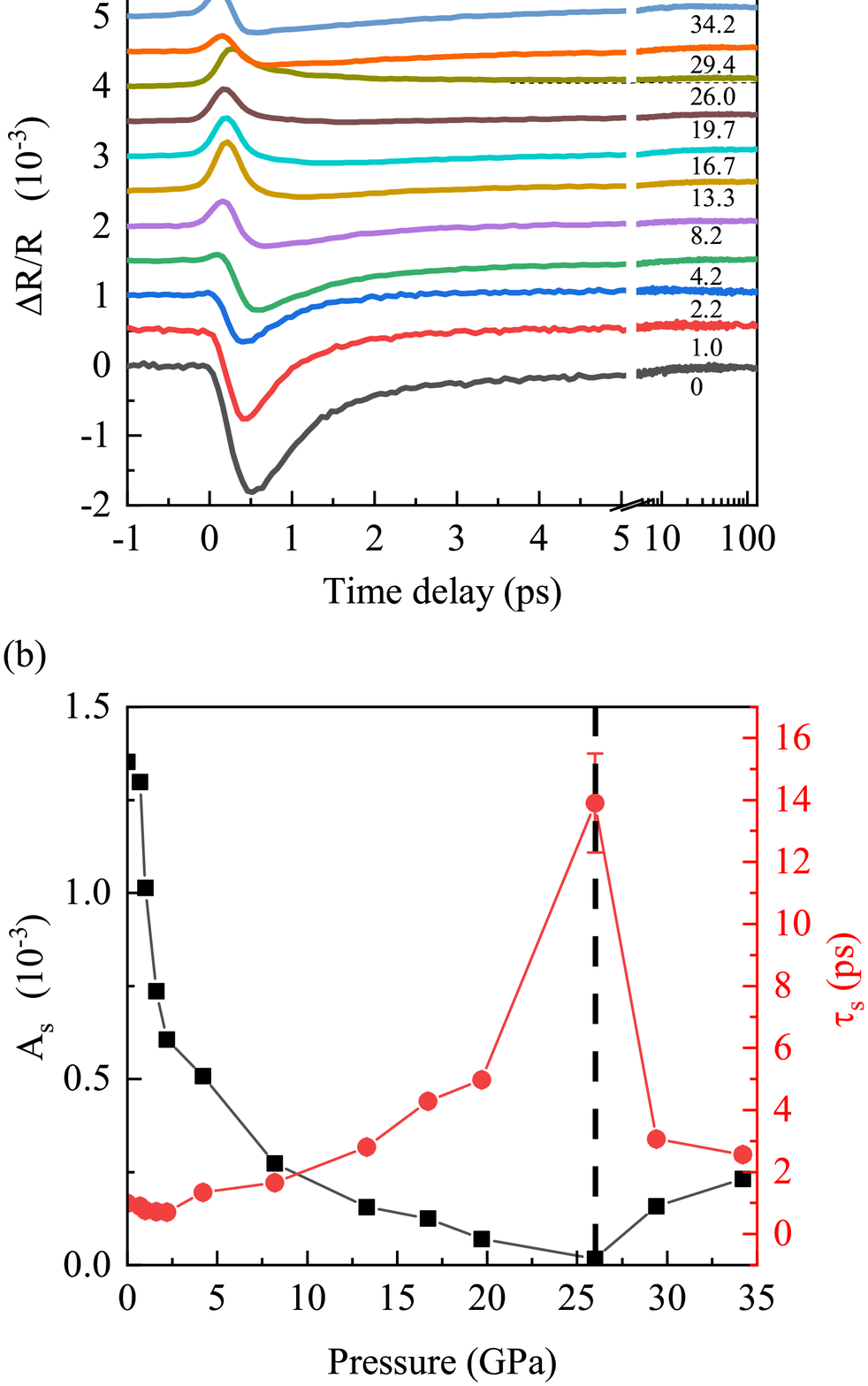}
	\subfigure{\label{3a}}
	\subfigure{\label{3b}}
	\caption{\textbf{Pump-probe spectra at 20 K and the extracted parameters as a function of pressure.} (a) Pump-probe spectra at various pressures at 20 K.  The dashed line indicates the existence of negative decay component at 26 GPa. (b) The extracted amplitude $\mathrm{A_s}$ and decay time $\tau_{\mathrm{s}}$ as a function of pressure. $\mathrm{A_s}$ decreases with increasing pressure before starts to increase above 26 GPa. $\tau_{\mathrm{s}}$ shows a quasi-divergent character, indicating the suppression of DW orders around 26 GPa.  Error bars are the standard error in the exponential fitting.}
	\label{3}
\end{figure*}

Assuming that $\Delta(T)$ follows BCS temperature dependence $\Delta(T) \approx \Delta(0) \tanh \left( 1.74 \sqrt{\frac{T_c}{T} - 1} \right)$, we fit $\mathrm{A_s}$ and $\tau_{\mathrm{s}}$ using Eq.(1) and (2).  The results, represented by the solid lines in Fig. \ref{1b}, yield a transition temperature $T_{\mathrm{DW}} \sim 151$ $\mathrm{K}$ and a gap energy $\Delta(0) \sim 66 \mathrm{~meV}$ which is in good agreement with the values previously reported by NMR\cite{kakoi2023multiband, dan2024spin} and optical conductivity spectroscopy\cite{liu2023electronic}. The excellent fit strongly supports our assumption of the formation of a gap in the electric density of states due to the development of DW order below $T_{\mathrm{DW}}$. We notice a similar work in Ref.\cite{li2024ultrafast} where no PB effect was observed at ambient pressure. This discrepancy is probably due to the inhomogeneous nature of La$_3$Ni$_2$O$_7$\cite{xie2024neutron,puphal2023unconventional,chen2024polymorphism,wang2024long,puphal2023unconventional,dong2023visualization,zhou2023evidence}, as discussed in Supplementary Note 2 and 3.

~\\
\noindent{\bf DW order evolution at high pressures}

To further investigate the evolution of DW order in La$_3$Ni$_2$O$_7$ as a function of pressure, we perform ultrafast pump-probe measurements under high pressure up to 34.2 GPa. Figure \ref{2} displays the temperature dependent transient reflectivity data at several selected pressures. The slow relaxation component with negative amplitude observed below $T_{\mathrm{DW}}$ persists across all pressures. The same fitting procedures described earlier were applied to the data under various pressures. The extracted parameter $\tau_{\mathrm{s}}$ is depicted in Fig. \ref{2} as scatter points. It is obvious that $\tau_{\mathrm{s}}$ diverges around $T_{\mathrm{DW}}$ for all pressures except 26 GPa. Above 29.4 GPa, the relaxation time $\tau_{\mathrm{s}}$ initially decreases slightly with increasing temperature, then sharply increases, exhibiting a quasi-divergent behavior at $T_{\mathrm{DW}} \sim 135$ K. This temperature dependence of $\tau_{\mathrm{s}}$ closely resembles that near ambient pressure, suggesting the re-opening of an energy gap under pressures above 29.4 GPa. 

In order to obtain more detailed information on the gap evolution, the $\Delta R / R$ signals as a function of pressure at 20 K are plotted in Fig. \ref{3a}. The negative amplitude monotonically reduces with increasing pressure and becomes indistinguishable at 26 GPa, above which the negative signal appears again. Figure \ref{3b} displays the fitting parameters $\mathrm{A_s}$ and $\tau_{\mathrm{s}}$ as a function of pressure at 20 K. As the pressure increases up to 2.2 GPa, $\mathrm{A_s}$ drops dramatically, accompanied by a slight decrease of $\tau_{\mathrm{s}}$. Upon further compression, $\mathrm{A_s}$ decreases gradually towards zero while $\tau_{\mathrm{s}}$ exhibits a quasi-divergence around 26 GPa. According to Eq.(2), the relaxation time increases with the decrease of the gap energy at fixed temperature and vice versa. Therefore, the observed increase in $\tau_{\mathrm{s}}$ with increasing pressure suggests a progressive suppression of the DW gap in this pressure range. Above 29.4 GPa, the increase of $\mathrm{A_s}$ and the decrease of $\tau_{\mathrm{s}}$ indicate the DW gap gets promoted again, consistent with the reappearance of the PB effect at higher $T_{\mathrm{DW}}$, as shown in Fig. \ref{2h}.

\begin{figure}[!htbp]
	\centering
	\includegraphics[width=3 in]{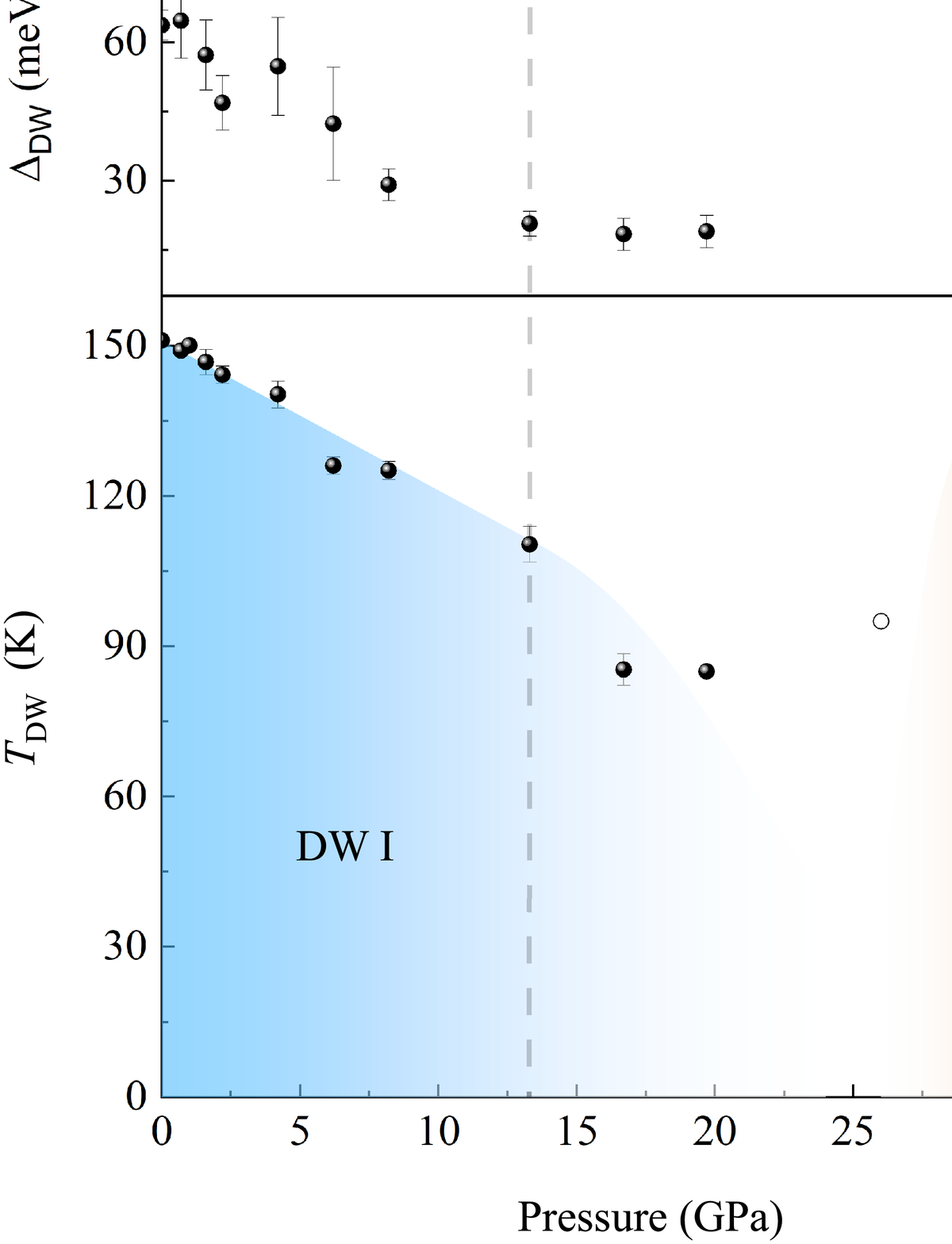}
	\caption{\textbf{Temperature-Pressure phase diagram of La$_3$Ni$_2$O$_7$ based on the pump-probe spectroscopy measurements.}  The upper and bottom panels show the extracted energy gap $\Delta_{DW}$ and the DW transition temperature $T_{DW}$, respectively.  The onset temperature of superconductivity $T_C$ obtained from resistance measurements\cite{sun2023signatures} were indicated as the starts  for comparison. The phase diagram is divided into two major regions, DW-I and DW-II, with a critical pressure of 26 GPa. Upon compression, the SDW order in the low region is significantly suppressed up to around 13.3 GPa as indicated by the shaded stripe,  before it completely disappears around 26 GPa. The phonon bottleneck effect at 26 GPa is too weak to extract a reasonable energy gap, hence a temperature of 100 K above which the negative decay component disappears is depicted as a hollow circle. Above 29.4 GPa, another DW-like order with a transition temperature of $\sim$ 135 K reemerges, which is probably related to the predicted CDW order. The error bars on the upper panel are calculated as the standard error in the RT model fitting of $A_s$ under pressure.}

	\label{4}
\end{figure}

The identical RT analysis was applied to the temperature dependence of the slow relaxation $\mathrm{A_s}$ and $\tau_{\mathrm{s}}$ at high pressures (Supplementary Note 4). The extracted $T_{\mathrm{DW}}$ values are summarized in the Temperature-Pressure phase diagram in Fig. \ref{4}. Based on the high pressure results above, the diagram can be divided into two major regions, DW-I, and DW-II, with a critical pressure of 26 GPa. In the low-pressure region, the DW transition is gradually suppressed from $ 151 \mathrm{~K}$ near ambient pressure to $ 110 \mathrm{~K}$ at 13.3 GPa. $T_{\mathrm{DW}}$ rapidly decreases to around 85 K at 16.7 GPa and then decreases slightly with pressure up to 19.7 GPa. Since the PB effect is too weak to distinguish at 26 GPa (Supplementary Note 4), a value of 100 K at which the negative decay component disappears, was added to Fig. \ref{4} as a hollow circle for comparison. Upon further compression, divergent behavior of $\tau_{\mathrm{s}}$ appears again near 135 K, suggesting the presence of another energy gap in the density of states. The transition temperature increases slightly with further increasing pressure. 

~\\
\noindent{\bf Discussion}

The present work provides clear evidences of the presence of DW orders in La$_3$Ni$_2$O$_7$ under high pressure. However, various broken-symmetry states such as superconductivity and spin/charge density wave will develop an energy gap below the phase transition temperature, resulting in a similar PB effect\cite{giannetti2016ultrafast,dong2023recent}. Combined with the recent RIXS\cite{zhou2024electronic}, $\mu$SR\cite{khasanov2024pressure,chen2023evidence} and NMR\cite{kakoi2023multiband, dan2024spin} experiments where an SDW transition around 150 K at ambient pressure has been identified, we attributed the DW order in the low-pressure region to the SDW. As shown in the upper panel of Fig. \ref{4}, the extracted gap amplitude decreases from approximately 66 meV near ambient pressure to around 20 meV at 13.3 GPa. The gradual suppression of the DW order with increasing pressure is consistent with the transport measurements\cite{PhysRevX.14.011040, zhang2023high,puphal2023unconventional,wu2001magnetic}.  It is worth mentioning that the PB effect above 13.3 GPa is relatively weak and the energy gap is roughly independent of pressure below 26 GPa, suggesting that the PB effect in the intermediate pressure range may have a different origin. Nevertheless, it does not originate from the superconductivity since the transition temperature obtained in present work is higher than the onset temperature of superconductivity in the resistance measurements\cite{sun2023signatures}, as indicated in Fig \ref{4}. Moreover, the  superconducting fraction volume has been demonstrated to be relatively low\cite{zhou2023evidence}, and hence superconductivity can not be captured by the ultrafast optical spectroscopy since it is a bulk-sensitive technique. After the long-range DW order is suppressed by pressure, the short-range order may persist in La$_3$Ni$_2$O$_7$\cite{PhysRevX.14.011040}, resembling cuprates\cite{wen2019observation} and iron-based superconductors\cite{David2010The}.  The short-range orders may induce the opening of a small gap in the density of states, as evidenced by the weak PB effect in the pressure range of 13.3 to 26 GPa. Another possibility is the coexistence of multiple structure variants in La$_3$Ni$_2$O$_7$ single crystals, including the majority La$_3$Ni$_2$O$_7$ (327) phase and minority La$_4$Ni$_3$O$_{10}$ (4310) and La$_2$NiO$_4$ (214) phase, as demonstrated by electron microscopy\cite{zhou2023evidence,wang2024bulk}. The SDW order below 13.3 GPa is unambiguous from the predominant 327 phase, while the weak features between 13.3 and 26 GPa may be contributed by the minor 4310 phase after the SDW in the majority 327 phase was suppressed by pressure above 13.3 GPa.  At 26 GPa, the PB effect is too weak to extract a reasonable gap, probably due to the complete suppression of the short-range orders in 327 phase and the DW order in 4310 phase\cite{ChinPhysLett.41.017401,zhang2023superconductivity}.

Recent theoretical works have indicated that electron-phonon coupling alone is insufficient to trigger superconductivity, suggesting that the Cooper pairing mechanism is unconventional in pressurized $\mathrm{La}_3 \mathrm{Ni}_2 \mathrm{O}_7$  and may originate from antiferromagnetic fluctuation\cite{gu2023effective, PhysRevB.108.L140505,PhysRevLett.131.236002,PhysRevLett.132.036502}. The suppression of SDW order above 13.3 GPa observed in the present work, coinciding with the onset of superconductivity in the transport measurements\cite{sun2023signatures,PhysRevX.14.011040}, suggests that magnetic fluctuations are particularly critical for understanding the pairing mechanism of superconductivity in this nickelate. Spin fluctuation has been considered to be the pairing mediator in unconventional superconductors, including cuprates\cite{armitage2010progress}, iron pnictides and chalcogenides\cite{dai2015antiferoomagnetic}, as well as infinite-layer nickelate\cite{paul2024spin}. Our phase diagram based on the ultrafast optical spectroscopic measurements, as shown in Fig. \ref{4}, indicates that the $\mathrm{La}_3 \mathrm{Ni}_2 \mathrm{O}_7$ share similarity with these superconductors in their paring mechanism.

First principle calculations revealed that $\mathrm{La_3Ni_2O_7}$ favors an antiferromagnetic ground state, under which the strong Fermi surface nesting evokes the electronic instability resulting in a potential structure transition from $Fmmm$ symmetry to $Cmmm$ or $Cmcm$ symmetry\cite{yi2024antiferromagnetic}. However, the authors in Ref.\cite{yi2024antiferromagnetic} also pointed out that the distortion of Ni-O bound length in the predicted CDW structure is less than 0.1\text {\AA}, making its experimental probe very challenging. The sensitivity of our technique to the presence of DW orders is further reinforced by the re-emergence of PB effect above 29.4 GPa. The linear temperature-dependent resistance above $T_c$, characteristic of strange-metal behavior, has been observed to persist up to 30 GPa\cite{zhang2023high, sun2023signatures}. The strange-metal behavior does not preclude the existence of DW-II order since the DW-like features in our ultrafast spectra are very clear even under pressure up to 13.3 GPa, while the resistance anomaly related to the DW order usually becomes undistinguishable above 3 GPa\cite{zhang2023high, sun2023signatures}.  Upon further compression, both the extracted transition temperature and energy gap increase slightly with increasing pressure, agreed nicely with the theoretic prediction\cite{yi2024antiferromagnetic}. Therefore, we attribute the DW-II phase to the predicted CDW.  Below $T_{\mathrm{DW}}$, the $\mathrm{A_s}$ follows a typical BCS-like temperature dependence, reflecting the behavior of the CDW order parameter (Supplementary Figure 4). Whether this CDW coexists or competes with superconductivity needs further investigations in the sample with a high superconducting volume fraction.

In summary, we have presented ultrafast optical pump-probe measurements on recently discovered nickelate superconductor $\mathrm{La}_3 \mathrm{Ni}_2 \mathrm{O}_7$ crystal under pressure up to 34.2 GPa.  By analyzing the data with RT model, the evolution of DW-like orders under high pressure is revealed and summarized in a phase diagram.  With increasing pressure, the SDW order, as demonstrated at ambient pressure, is significantly suppressed up to 13.3 GPa before it completely disappears around 26 GPa. Intriguingly, at pressures above 29.4 GPa, another DW-like order with a transition temperature of approximately 135 K re-emerges, which is probably related to the predicted CDW order. Our results not only provide the experimental evidence of the DW evolution under high pressure, but also offer insight into the underlying correlation between the DW order and superconductivity in pressured $\mathrm{La}_3 \mathrm{Ni}_2 \mathrm{O}_7$.

~\\
\noindent{\bf \large Methods}

\noindent{\bf  Sample growth and characterization.} Single-crystalline La$_3$Ni$_2$O$_7$ samples were grown using a vertical optical-image floating-zone method at an oxygen pressure of 15 bar and a 5-kW Xenon arc lamp (100-bar Model HKZ, SciDre)\cite{sun2023signatures,liu2022evidence}. A small piece of sample was cut from the crystal to measure the resistivity using standard four-probe method. The sample exhibits metallic behavior at ambient pressure and undergoes a clear drop in resistance around 80 K at 16.7 GPa (Supplementary Note 5). Crystal structures of the samples were investigated by x-ray diffraction (XRD, Empyrean, Cu target) at 300 K. The results indicate that the sample is a bilayer structure in $Cmcm$ space group at room temperature and ambient pressure (Supplementary Note 6). The sample with flat surface cut from the same crystal was used for the transient reflectivity measurements.
~\\

\noindent{\bf  Measurements under high pressure and low temperature.} High pressure was generated by screw-pressure-type nonmagnetic Be–Cu alloy diamond anvil cell (DAC) with a 500 $\mu \mathrm{m}$   culet. Fine KBr powders were used as the pressure transmitting medium, which has been demonstrated could offer quasi-hydrostatic pressure condition for the pump probe measurments\cite{PhysRevB.109.064307}. The sample chamber with a diameter of 300 $\mu \mathrm{m}$ was made in a Rhenium gasket. $\mathrm{La_3Ni_2O_7}$ crystal with size of 150 $\mu \mathrm{m}$ was loaded in the center of the chamber and a small ruby ball was placed aside the sample. The DAC was loaded in a continuous flow liquid helium cryostat with temperature varying from 10 to 300 K. An additional thermal sensor was mounted on the force plate of the DAC for precise measurement of sample temperature. The pressure was calibrated using the ruby fluorescence shift at low temperatures for all the pump-probe experiments.

~\\
\noindent{\bf Pump-probe measurement.} An achromatic pump-probe system based on a mode-locked Yb:KGW laser system was employed. The laser pulses with wavelength of 800 nm and repetition rate of 50 kHz was generated by the optical parametric amplifier, which was divided into two beams. One served as the probe beam, and another passed through a BBO crystal to generate a 400 nm pump pulses. The pump and probe beams were focused onto the sample surface through a 5$\times$ objective lens. The focused spot diameters of the pump and the probe pulse were 37 and 17 $\mu$m, respectively. The pulse duration, after passing through the cryostat window and diamond anvil, was measured to be 50 fs. In the temperature and pressure dependence measurements, the pump and probe fluences on the sample were kept at 45 and 9 $\mu \mathrm{J} / \mathrm{cm}^{-2}$, respectively. The pump beam was modulated by a chopper with a frequency of 433 Hz and the reflected pump beam was filtered out. The reflected probe beams traversed through the same objective lens, received by a photo-diode detector and sampled by a lock-in amplifier to enhance the signal-to-noise ratio. The relative change of reflectivity $\Delta R(t) / R_0=\left[R(t)-R_0\right] / R_0$, where $R$ and $R_0$ are the reflectivity of the probe with and without the presence of pump pulses, respectively, was recorded as a function of the time delay between the pump and probe pulses.

	

\begin{acknowledgments}
	This work was supported by the National Natural Science Foundation of China (Grants No. 11974414 (X.B.W.), No. 12374050 (F.H.), No. 12134018 (J.L.L.), No. 12425404, and No. 12174454), the National Key Research and Development Program of China (Grants No. 2023YFA1608900, No. 2021YFA1400300, No. 2021YFA0718700, No. 2023YFA1406000, No. 2023YFA1406500), the GuangDong Basic and Applied Basic Research Foundation (Grants No. 2024A1515030030 and No. 2024B1515020040), the Shenzhen Science and Technology Program (Grant No. RCYX20231211090245050), and the Guangdong Provincial Key Laboratory of Magnetoelectric Physics and Devices (Grant No. 2022B1212010008).  This work was carried out at the Synergetic Extreme Condition User Facility (SECUF).

\end{acknowledgments}




~\\

	\clearpage

\end{document}